\documentclass[aps,pre,twocolumn,floatfix,longbibliography,superscriptaddress]{revtex4-1} 

\AtBeginDocument{%
    \newwrite\bibnotes
    \def\bibnotesext{Notes.bib}
    \immediate\openout\bibnotes=\jobname\bibnotesext
    \immediate\write\bibnotes{@CONTROL{REVTEX41Control}}
    \immediate\write\bibnotes{@CONTROL{%
    apsrev41Control,author="08",editor="1",pages="1",title="0",year="1"}}
     \if@filesw
     \immediate\write\@auxout{\string\citation{apsrev41Control}}%
    \fi
}%

\usepackage[normalem]{ulem}

\usepackage{amsmath,amssymb,amsfonts}
\usepackage{graphicx} 
\usepackage{times}
\usepackage{bm, mathtools}
\usepackage{color}
\usepackage[dvipsnames]{xcolor}

\definecolor{linkcolor}{rgb}{0,0,0.66} 
\usepackage[colorlinks=true, pdfstartview=FitV, linkcolor= linkcolor, citecolor= linkcolor, urlcolor= linkcolor, hyperindex=true,hyperfigures=true]{hyperref} 

\renewcommand{\cite}{\citep}
\renewcommand{\vec}{\bm}

\newcommand{\UU}{{\Delta E_{12}^S}}
\newcommand{\UUU}{{\Delta E_{13}^S}}

\begin{document}

\title{Aging of living polymer networks: a model with patchy particles}

\author{Stefano Iubini}
\affiliation{Consiglio Nazionale delle Ricerche, Istituto dei Sistemi Complessi, via Madonna del Piano 10, I-50019 Sesto Fiorentino, Italy}
\affiliation{Dipartimento di Fisica e Astronomia, Universit\`a di Padova, Via Marzolo 8, I-35131 Padova, Italy}
\author{Marco Baiesi}
\affiliation{Dipartimento di Fisica e Astronomia, Universit\`a di Padova, Via Marzolo 8, I-35131 Padova, Italy}
\affiliation{Sezione INFN di Padova, Via Marzolo 8, I-35131 Padova, Italy}
\author{Enzo Orlandini}
\affiliation{Dipartimento di Fisica e Astronomia, Universit\`a di Padova, Via Marzolo 8, I-35131 Padova, Italy}
\affiliation{Sezione INFN di Padova, Via Marzolo 8, I-35131 Padova, Italy}

\begin{abstract}
Microrheology experiments show that viscoelastic media composed by wormlike micellar networks display complex relaxations lasting seconds even at the scale of micrometers. By mapping a model of patchy colloids with 
suitable mesoscopic elementary motifs to a system of  worm-like micelles, we are able to simulate its relaxation dynamics, upon a thermal quench, spanning many decades, from microseconds up to tens of seconds. 
After mapping the  model to real units and to experimental 
scission energies, we show that the relaxation process develops through a sequence 
of non-local and energetically challenging arrangements.
These adjustments remove undesired structures formed as a 
temporary energetic solution for stabilizing the thermodynamically unstable free caps of the network.
We claim that the observed scale-free nature of this stagnant process may complicate 
the correct quantification of experimentally relevant time scales as the Weissenberg number.
\end{abstract}

\maketitle

\section{Introduction}
Many soft materials with important industrial and medical applications
are formed by thermodynamic self-assembly of elementary constituents
dispersed in aqueous solution~\cite{wei06}. This process gives rise to
networks of linear or branched fibers or even more complex objects~\cite{jehser2020}
  that can continuously break,
rejoin or get entangled. These ``living'' materials~\cite{cat06} have remarkable
viscoelastic properties that are typically studied either by imposing
a shear stress with a rheometer~\cite{wei06,lar99}
or by dragging a micro-sized probe through the
medium as in active microrheology~\cite{gom15,ber18}. The response and
relaxation dynamics of living polymers are the result of large-scale
reorientation and mutual reptation of the fibers that, in turn,
depend on small scales mechanisms such as scission and rewiring of
branches. This multiscale process can give rise to global relaxation
dynamics that, for systems as wormlike micelles, has been measured to be
of the order of seconds~\cite{gom15,ber18}.

An important issue is the relation between the macroscopic dynamical
response of the system, measured experimentally,
and its less accessible local structural rearrangements. From the theoretical
side, mean-field theories and microscopic constitutive models have been
successfully developed~\cite{tur90,dry92,gra92,tur93,cat06}
to rationalize the outcomes of rheological experiments.
However, these studies often dealt with small temperature jumps, which may not
lead to the significant disturbance of the micellar networks
as, e.g., in microrheological experiments  at large Weissenberg numbers where 
a microbead is moved in the medium at high velocities~\cite{dea10,poo12}.
At the same time, numerical studies~\cite{pad05,pad08,hug11,dha15,wan17,man18},
performed at a coarse-grained molecular level on specific
systems (cationic cylindrical micelles and cetyltrimethylammonium
chloride micelles) have
either looked at the competition between shear flow and internal
network dynamics or estimated the strands scission and branching free
energies of these systems. However, these  models, being still quite
detailed and system-specific, cannot have access to the scales
spanned by rheological experiments (of the order of seconds and above
$\mu m^3$) or address universal features of the phenomenon.

%
\begin{figure*}[!t]
\centering
\includegraphics[width=0.95\textwidth]{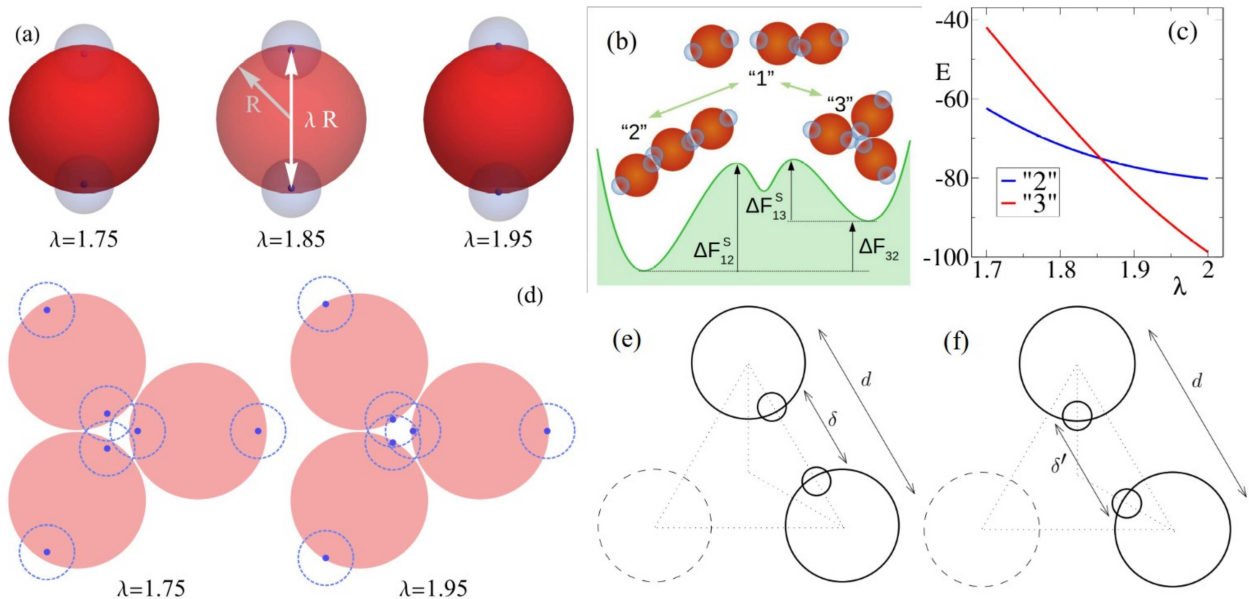}
\caption{
  (a)
  Example of three patchy particles: Two sticky spots of range $\sigma$
  (light blue spheres of radius $\sigma$ with darker dot at their center)
  form a rigid body with a (red) repulsive core of radius $R$.
  The sticky spot centers are at distance $\lambda R$ from each other.
  (b)
  Sketch of a possible free energy landscape for a trimer, i.e.~a system with three patchy particles/micellar units.
  (c)
  Ground-state energy of linear-trimer (``2'' state, blue) and  branching-point (``3'' state, red) configurations as a function of the branching parameter $\lambda$.
  (d) Sketch of a triple contact for $\lambda=1.75$ (sticky spots farther than the interaction range $\sigma$) and for $\lambda=1.95$ (sticky spots closer than $\sigma$).
  (e) and (f):
  Geometrical details of the (planar) linear and branched configurations.
   The total potential  energy of a strand is $E_l=2 u_l$ due to a full overlap
of the sticky spots  while, in a branching point, the same quantity is
$E_b=3 u_b$ with a bond energy $|u_b|<|u_l|$
which tends  to $u_t$ only when $\lambda$ is sufficiently
large.
}
\label{fig:sk}
\end{figure*}

A generic understanding of the transient behavior of living polymers
can be sought by addressing the following questions: which are the
dominant local structures (or ``microstates'') that compete
dynamically in the rearrangement of a relaxing living random network?
How does the long-time relaxation depend on the mechanisms of local
rewiring and scission of the fibers?  Is the time dependence of
relevant observables undergoing relaxation characterized by simple
exponentials or is it more similar to a dynamical scaling with power
laws?  Simulations of coarse grained models at mesoscopic level are an ideal tool for
answering these questions: they allow a direct investigation both at
short/small and long/large scales and they may adopt a well controlled
protocol for driving the system out of equilibrium.

In this paper, we introduce a simple, coarse-grained,  description
of a thermally driven self-assembly process generating a
living network of fibers by means of the merging, scission and rewiring of
basic units that represent a portion of the tubular structure.
These units are described by patchy particles~\cite{sci09,li20,audus2018,russo2011,russo2011b,rovigatti2013,teixeira2017,rov18,chen11,rom11,rom11b,rom12,mah16,rei16,esl19}.

By using large-scale Brownian dynamic simulations and analytical
arguments to estimate the free-energy barriers between the
competing ``microstates'', we characterize the underlying microscopic
dynamics that leads to the global relaxation towards the formation of the
random networks.  To go beyond the regime of small perturbations a wide temperature quench
is imposed on the system.

The paper is organized as follows. In section~\ref{sec:mod} we introduce the coarse-grained model
and the numerical set-up.
 In section~\ref{sec:res} we analyze the thermodynamic and kinetic properties of local motifs and 
their effect on  the large-scale relaxation dynamics of the system upon a temperature quench towards a self-assembled equilibrium state.  
Section~\ref{sec:disc} is devoted to a final discussion of the results and to highlight future
developments  that can be explored within this approach.
Finally, the appendix contains the details of the numerical methods employed for this study.

\section{Model}
\label{sec:mod}

The coarse grained model we consider  belongs to the vast class of patchy particle models 
used in literature to study self-assembled structures~\cite{sci09} but it is specifically designed 
to describe living polymer networks and particularly worm-like micelles.
Let us first introduce its basic properties in 
the context of patchy colloids (subsection~\ref{ssec:coll}) and later discuss a more specific
 connection with micellar systems (subsection~\ref{ssec:mic}).

  \subsection{Patchy colloids}\label{ssec:coll}
The elementary unit of our coarse-grained description is a rigid body consisting of 
three sub-units: there is a central  core of radius $R$ (red spheres in
Fig.~\ref{fig:sk}(a)) and two radially opposite (i.e. antipodal) ``patches" or ``sticky spots'' (blue
transparent spheres) fixed at distance $\lambda R$. 
The effective interaction between units is the sum of two contributions: a steric (excluded volume) 
interaction between the core beads and a mutual attraction between patches belonging to different rigid bodies. The first is accounted for a shifted and truncated Lennard-Jones potential
with characteristic amplitude $\epsilon$ and range $R$. Patch-patch attraction, that would lead to the aggregation of units at sufficiently low temperatures, is instead described by a
truncated Gaussian potential with range $\sigma= 0.4 R$ and amplitude $A=40 \epsilon$ (this amplitude
allows to recover free energies at $T\approx 1$  that are comparable to the
ones measured in  micellar systems),  see Appendix~\ref{sec:app} for more details.

Two main characterizing aspects of our model are worth noticing.
First,  in most of the patchy colloids models~\cite{sci09} the branching points
of  the assembled structures are generated either by  introducing a
fraction of colloids decorated by three or more attractive
patches~\cite{li20,audus2018}
or by allowing for different types of patch-patch interactions~\cite{russo2011,russo2011b,rovigatti2013,teixeira2017}.
In either case, a better control of the phase
diagram is achieved by satisfying a single-bond-per-patch condition~\cite{rov18}.
In our model  only bifunctional patchy beads are considered and the local branching 
emerges naturally from the formation of multiple contacts between patchy particles.
Note that similar systems were studied experimentally~\cite{chen11}
and numerically~\cite{rom11,rom11b,rom12} (see also some more recent work~\cite{mah16,rei16,esl19}).
The second aspect concerns  the presence of the tunable {\em branching parameter} $\lambda$.
This has been introduced to mimick the propensity of worm-like micelles to branch as a function 
of salt concentration (see section~\ref{ssec:mic}). In particular, by varying $\lambda\leq 2$  
($\lambda=2$ corresponds to standard surface interactions between patches),
one can change the depth of the  sticky spots within the repulsive core
and hence bias the system towards the formation of either linear strands or branches (see Fig.~\ref{fig:sk}(b) for the simplest case of trimers).  
More precisely, for large values of  $\lambda$ the units tend to form  ``vertex'' or
``3 contact''  states (see Fig.~\ref{fig:sk}(b)) rather than the ``strands'' or  ``2 contact'' states.

This behavior can be explained quantitatively by minimizing analytically the total
potential energy of the vertex and strand configurations of a trimer, see Fig.~\ref{fig:sk}(c). 
As a matter of fact, we find that the formation of strands is
energetically advantageous with  respect to branch points only for sufficiently
small values of  $\lambda$. Branch points become instead energetically more convenient if $\lambda$ gets close enough to $2$. For example, in the two situations represented in Fig.~\ref{fig:sk}(d), for $\lambda=1.75$ the mutual attraction between sticky spots take place at a distance larger than the interaction range $\sigma$, reducing the stability of the trimer, while  for $\lambda=1.95$ the three pairs of sticky spots are all at a distance within $\sigma$ and thus very stable.

The effect of the branching parameter $\lambda$ can be further rationalized as follows.
Let us focus on a linear dimer (see Fig.~\ref{fig:sk}(e)) in its mechanical equilibrium position with bond energy $u_l<0$ determined by the 
distance $d$ between unit centers and the distance $\delta=d-\lambda$ between sticky centers. 
The branched configuration can be obtained by rotating the two reference units by
angles of $30^\circ$ with respect to their centers on the plain containing the trimer  (Fig.~\ref{fig:sk}(f)). In this rotated configuration the two sticky spots 
increase their distance $\delta'>\delta$ thus decreasing their attracting force, while the repulsive centers of the units are not
modified significantly. Since $ \delta' > \delta$ the single bond energy $u_b>u_l$.
However, due to the different number of bonds in branched and linear trimers,
the overall energy of the branched trimer $E_b=3u_b$ can be lower and thus more stable than the linear trimer configuration with  energy $E_l=2u_l$. This takes place if $\lambda$ is large enough to keep $\delta'$ within the range of attraction of the sticky spots.
Upon decreasing $\lambda$, one meets a situation in which $\delta'$ is longer than the range of attraction between sticky spots and hence the branched configuration becomes less stable than the linear one.

\subsection{Wormlike micelles}\label{ssec:mic}
The patchy colloid model described above is rather generic  but it can be specialised on  
worm-like micelles for which a wealth of experimental
data is nowadays available. In this context, it is known that the
spatial organization of micelles in spherical, linear or branched
structures depends on several parameters   such as salt
concentration~\cite{cat06,dha15} or the spontaneous curvature of
surfactants~\cite{tlusty2000,zilman2004}.  This feature makes
large-scale numerical simulations  rather challenging,  even at
coarse-grained level, since a detailed description of micelles
recombination dynamics requires to resolve the mutual interaction of
solvent, salt and surfactant particles at the subnanometric length
scales~\cite{dha15,man18}.  In contrast, in our minimal model
the effect of ionic particles and curvature is taken into
account \emph{effectively}  through the tunable branching parameter $\lambda$.

%
\begin{figure}[!t]
\centering
\includegraphics[width=0.56\columnwidth]{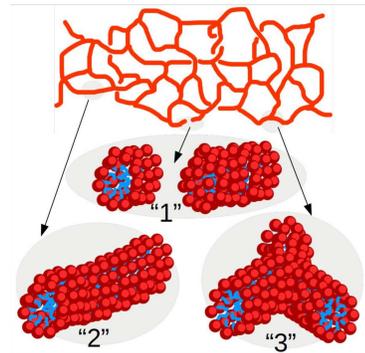}
\caption{Sketch of a wormlike micellar network and of selected
  portions where each surfactant is depicted as a hydrophilic
  (red) head  and a hydrophobic (blue) tail.
  The numbering of the network portions follows that of the 
   corresponding patchy particle states of Fig.~\ref{fig:sk}(b).
  }
\label{fig:sk2}
\end{figure}

In this approach an ensemble of surfactants forming a globular micelle or the  segment
of a wormlike network, as
those sketched in Fig.~\ref{fig:sk2}, is mapped to a single
 {\em  micellar unit}.
In this context, one may think of $\lambda$ as a parameter describing the strength of  salt concentration. 
 This is indeed known to shrink the effective size of the hydrophilic heads of the surfactants, thus favoring the appearance of structures with negative 
surface curvature as hubs in the micellar network.

Note that the resolution of our coarse grained model cannot describe effects below the size of a globular micelle. In particular, unlike previous models~\cite{pad05,pad08,hug11,dha15,wan17,man18}, the dynamics of a single surfactant is not included in our description. 
On the other hand, the fact that the basic element of our model 
is not representing a single surfactant but rather a collection (i.e. some dozens) of them, should not affect the large scale dynamics of the network structure that we aim to explore here.

\subsection{Simulations}
The dynamics of the system follows a set of coupled Langevin
equations for the interacting units
in an implicit solvent at fixed volume and constant temperature $T$
(henceforth measured in units of $\epsilon$,
with Boltzmann constant $k_B=1$ and $\epsilon=1$).
Numerical simulations are performed with LAMMPS engine~\cite{Plimpton1995a}.
For a more detailed description of the model
and for a mapping of the simulation units to those
of a wormlike micellar system, see the appendix.

By assuming, for instance,  that the core radius of the elementary
unit length is of the order of the average width measured in wormlike
micelles, i.e.~$R\approx 5 n m$, the largest box considered in our simulations with  dimensions
 $108 R\times 108 R \times 324 R$  maps to a system volume  $\approx 0.5 \mu m^3$.
Moreover, our runs simulate systems of up to $N=6000$ units; this would correspond 
to a  number of molecules of surfactants of the order of $10^5$~\cite{hyde1996}.
Finally, since the simulation time unit $\tau$ is mapped to $2\mu s$ in the real system (see appendix),
the longest simulation time we can reach, $t=10^7 \tau$, corresponds approximately to $20\,s$
and is comparable with the duration of recent experiments~\cite{ber18,zou19}.
Simulations are performed with an integration time step equal to $10^{-2} \tau$.
We have verified that this time step suffices to ensure suitable numerical accuracy in 
all regimes considered in this study.

\begin{figure*}[tb!]
\centering
\includegraphics[width=0.9\textwidth]{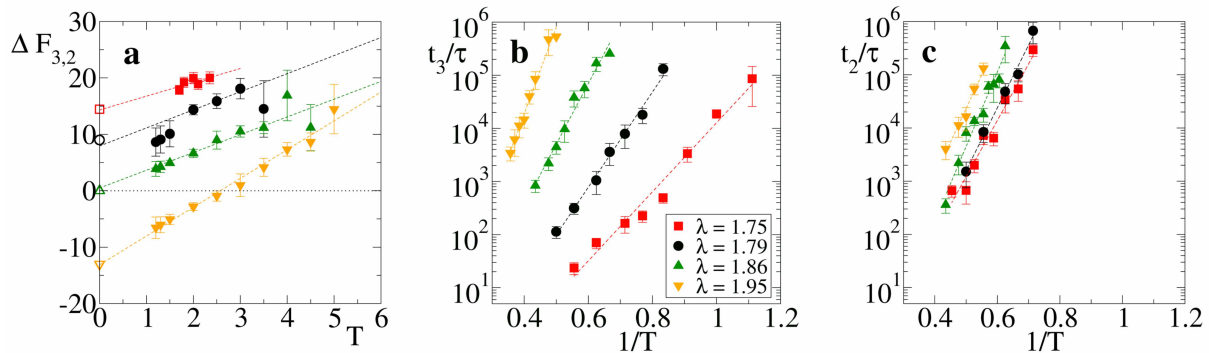}
\caption{Thermodynamic and kinetic properties of a trimer.
  (a) Free energy difference $\Delta F_{32}$ vs.~temperature $T$ for $\lambda=1.75$, $1.79$, $1.86$ and $1.95$.
  For the same values, $-\Delta S_{32}=2.4$, $3.2$, $3.2$ and $5.1$ from linear fits, see dashed lines.
Open symbols reported on the vertical axis correspond to the energy difference from Fig.~\ref{fig:sk}(c), at
$T = 0$.
(b) Average scission time $t_{3}$ for a trimer initially arranged in the configuration ``3'' and (c) average scission time $t_{2}$ for a strand ``2'', as a function of the inverse temperature $1/T$.  
Dashed lines are Arrhenius fits of $\UUU$ and $\UU$.
 In all panels different symbols refer to different values of the branching parameter $\lambda$ (see legend in panel (b))
}
\label{fig:trimer}
\end{figure*}

We describe the time evolution of the  network 	 by monitoring the formation/disappearance of 
its local motifs, such as dangling ends and branched structures.
These are classified by looking at the number of contacts shared by their patches. 
A sticky spot is defined to share a contact with another one if the distance between their centers is $\le R$, see appendix for further details. A dangling end contains a unit with $k=1$ contacts, while each branch ``3'' contains three units each with $k=3$ contacts, when they are part of a strand in the network (that is, a chain of elementary units with $k=2$) which ends either with dangling ends or branching points.
Finally, we define the length $L$ of a strand as the total number of linear bonds  between its extrema. 
For example, a dimer composed of two units with $k=1$ has length $L=1$, while a sequence on five units with $k=2$ between two joints has length $L=6$.
We assign $L=0$ to isolated units ($k=0$).

\section{Results}\label{sec:res}

\subsection{Thermodynamics and scission time of local motifs}\label{ssec:loc_th}

We first analyze the equilibrium
thermodynamics of the smallest motifs assembled  in the network, namely
the strands and the branch points (respectively state ``2'' and ``3''
in Fig.~\ref{fig:sk}(b)). 
In presence of thermal fluctuations, the zero-temperature scenario of Fig.~\ref{fig:sk}(c) is
modified by entropic contributions and a comparison between  the free energies $F$ of the
microstates ``2'' and ``3''  must be performed.
Note that in micellar solutions thermal fluctuations not only determine the configurational entropy of the motifs,
but also modify  the typical energies of strand and branch points~\cite{zilman2004}. 
At our coarse-graining level, this would result to a branching parameter $\lambda$ that depends also on  temperature. In this study, however, since we are interested in identifying  the 
role of purely entropic contributions  $T$ and $\lambda$ are tuned  independently. 
On the other hand, once the model is fully characterized, any mapping to a given experimental 
condition  of salt concentration and temperature can be obtained  through a suitable 
selection of the $\lambda$ and $T$ values.

To estimate the free
energy difference  $\Delta F_{32} = F_3 - F_2$ we have carried out Langevin
simulations with parallel tempering sampling
technique~\cite{Swendsen1986,tesi1996}  on a system with three units
confined within a small fixed volume (see appendix for details).  In
Fig.~\ref{fig:trimer}(a) we report $\Delta F_{32}$ as a function of
$T$ for four values of $\lambda$. It displays a linear growth
with $T$, with an intercept at $T=0$ that matches quite well the
 energy differences found analytically from Fig.~\ref{fig:sk}(c), see open symbols.
 Entropy differences $\Delta S_{32}$ are thereby identified by the (negative) slope $\partial
\Delta F_{32} / \partial T$, see  dashed lines.  Note that the
combination of the negative entropy gap $\Delta S_{32}$ with a
$\lambda$-dependent energy potential may result  in a situation where the
most favorable thermodynamic state is $T-$dependent, as for
$\lambda=1.95$, where branch points are stable only at low $T$.
Analogous results involving temperature-dependent effective valence were previously found in
particles models with explicit dissimilar patches~\cite{russo2011b}.
These findings match the prediction~\cite{man18} that  wormlike
micelles may form networks with a variety of features, depending on
the inclination of their tubular structures to branch.

To see how local thermodynamic properties affect the dynamics of
network formation, we have computed the average  scission time either from the
strand ``2" or from the branch point ``3" to the detached
dimer-plus-monomer state ``1", see Fig.~\ref{fig:sk}(b).
Scission times have been computed from equilibrium simulations for different temperatures $T$ 
and different values of the branching parameter $\lambda$, see appendix for details. 
  They follow a standard Arrhenius law once reported as a function of $1/T$.  For branched initial configurations
(Fig.~\ref{fig:trimer}(b)), we find average scission times
$t_3 \sim \exp(\UUU/T)$ with energy $\UUU$  whose fitted values are
$\UUU\approx 15$, $21$, $26$, $37$ (in units of $k_B T$), respectively
for $\lambda=1.75$, $1.79$, $1.86$, and $1.95$.  Similarly, the
scission times for strand initial configurations is $t_2  \sim \exp(\UU/T)$
(Fig.~\ref{fig:trimer}(c)) and the fitted
values of the scission energies $\UU$ are $24$, $27$, $32$, and
$29$ for the same $\lambda$ values.  Importantly,
these values of $\UU$ and $\UUU$ are in a range
$\approx 15\div 35\,k_B T$ fully compatible  with the current
estimates of the scission energies (or enthalpies) of micellar
systems~\cite{vog17,jia18,man18,zou19}.

\subsection{Relaxation dynamics}

\begin{figure}[!t]
\centering
\includegraphics[width=0.47\textwidth]{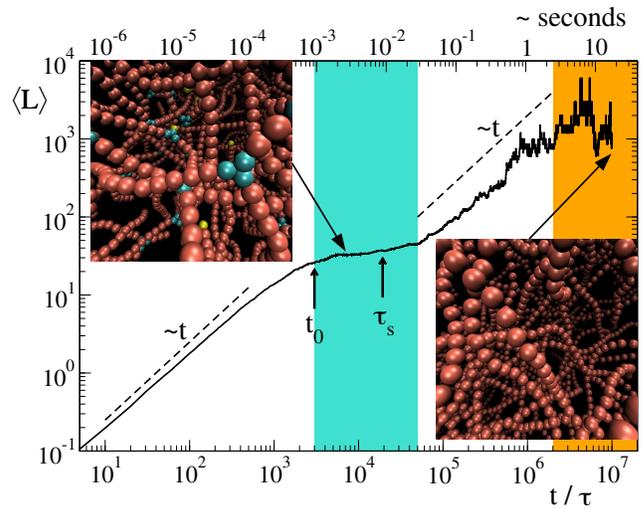} 
\caption{Evolution of the average strand length $\langle L \rangle$ for a system of $N=6000$ particles with $\lambda=1.75$, and $T=1$ (thick black curve). The cyan band marks the metastable regime with excess of branches (see an example in the upper inset) while the orange band identifies the regime of thermodynamic equilibrium (lower inset). 
  Vertical arrows indicate the condensation time scale $t_0\simeq 3\times 10^3 \tau\approx 1$ms and the typical time $\tau_s\simeq 2\times 10^4 \tau$ of network rearrangements.
  In the two configurations of the network, units with $k=1,2,3$ contacts are drawn, respectively, in yellow, red and cyan.
} 
\label{fig:avL}
\end{figure}

Once clarified the  equilibrium and kinetic properties of the basic units 
of the  model, we now turn to the main topic of this work, namely the study of the 
long time relaxation process of the system towards  relatively large networks  at equilibrium.
This is carried out  by large-scale simulations in which the system starts from a solution of freely diffusing globular 
micelles (which in our model are represented by isolated units and initialised with $T\gg 1$).
At sufficiently low $T$, isolated units start to aggregate into intricate networks, 
see the inset of Figs.~\ref{fig:avL} and~\ref{fig:lambda2}(b).
Their typical structures  bear a strong resemblance of those appearing on smaller scales 
in less coarse-grained models~\cite{pad05,pad08,hug11,dha15,wan17,man18}.
In Fig.~\ref{fig:avL} we report the time evolution of the mean strand length
$\langle L \rangle $  for $\lambda=1.75$, $T=1$ and $N=6000$,  where the symbol $\langle \cdot \rangle$ indicates
the average over all strands found in the system at time $t$. 
Note that, for the chosen values of $\lambda$ and $T$, equilibrium thermodynamics predicts a large predominance
of strands over branching points, see red squares in Fig.~\ref{fig:trimer}(a). Hence the system is expected
to arrange itself in long strands.
However,  after a relatively fast power-law growth for a time $t_0\sim 10^3 \tau$, $\langle L
\rangle $ reaches first a plateau that lasts for a decade (see cyan band, whose width is $\tau_s\sim10^4\tau$)
and then slowly increases as a second power law towards the expected
equilibrium value (orange band).
From numerical fits, we have found that the exponent of the two power-laws
 is $\simeq 1$, see dashed lines in Fig.~\ref{fig:avL}.  
The  intermediate plateau on $\tau_s$  indicates 
that the growth 
of the network cannot be described in terms of a simple coalescence
process~\cite{chandra1943,carnevale1990}, due to the presence of a metastable state
characterized by an excess of branching points (see
top-left inset in Fig.~\ref{fig:avL}). As a result,  longer strands can be
progressively assembled by slow rearrangements of the whole structure.


The macroscopic relaxation process  is
further understood by looking at the dynamical evolution of units
classified by their contacts.  Let $N_k$ be the number of units with
$k$ contacts and $f_k = N_k / N$ their fraction.   Due to the core
repulsion, for $\lambda \lesssim 2$ a unit may establish at most $k=4$
contacts (two per side).

\begin{figure}[!tb]
\centering
\includegraphics[width=0.99\columnwidth]{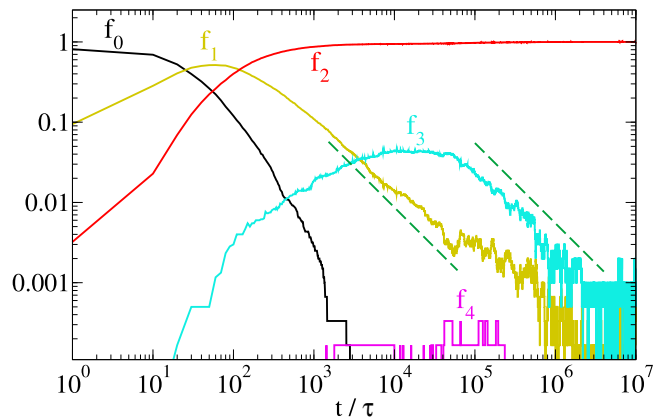}
\caption{Evolution of the relative density $f_k$ of units with $k$-contacts for the same 
relaxation dynamics of Fig.~\ref{fig:avL}
($N=6000$, $\lambda= 1.75$ and $T=1$).
  Dashed lines represent a scaling $\sim t^{-1}$.
}
\label{fig:cont}
\end{figure}

In Fig.~\ref{fig:cont} we plot the fractions $f_k$ vs time for the
same system of Fig.~\ref{fig:avL}. After a quick condensation to a
network ($f_0\to 0$ for $t\to t_0$), the
network does not contain the equilibrium proportion of network motifs.
The fraction $f_1$ of units at dangling ends tends slowly to zero,
with a scaling compatible with a power law $f_1 \sim
t^{-1}$. Eventually, at $t_{\rm eq}\approx 2\times 10^6 \tau$, mapped
to $ \approx 10 s$, the fraction $f_1$ reaches its asymptotic value.
This very slow reduction of the number of strands with dangling ends
highlights a stagnant global relaxation of the network.  Also the
fraction $f_3$ of units meeting at branches slowly decreases as
$f_3\sim t^{-1}$ till $t_{\rm eq}$ and then
fluctuates around its equilibrium value. 
We can approximate the asymptotic ratio $\bar{f}_3/\bar{f}_2\simeq \bar{f}_3$ (the bar over $f_k$ denotes the long-time value of $f_k$) by the
equilibrium expression $\bar{f}_3/\bar{f}_2=\exp(-\Delta f_{32}/T)$,  where
$\Delta f_{32}=\Delta F_{32}/3$ is  the free energy difference per unit 
derived from the single trimer problem, see section~\ref{ssec:loc_th}.
For $\lambda=1.75$ and $T=1$, this gives $\bar{f}_3/\bar{f}_2\simeq 0.003$.
This prediction, however, is not fully quantitative, as it neglects 
the role of trimer interactions (and even more complex many-body interactions) which are present in the macroscopic
system.
Finally, a very small amount of units with four contacts ($k=4$) is temporarily 
observed during the relaxation process, in correspondence of the
maximum concentration of branches $f_3$. This is not surprising, as four-contacts states
arise when the branched geometry affects both sides of a unit. Since their energetic cost
is essentially twice the cost of a single branch, their asymptotic concentration 
is expected to be of the order of $(\bar{f}_3)^2$, a behaviour which, however, was not observed 
in our simulations due to finite size and finite time effects.

The slow scalings  of $f_1$, $f_3$, and $\langle L\rangle$ suggest
that the dynamics within the network proceeds through a sequence of
frustrated rearrangements. For instance, a dangling end joining its
free cap to the side  of a strand would form a branch. This local
state is however thermodynamically less favorable (see Fig.~\ref{fig:sk}(b)),
 thus leading to a
subsequent decay back to a dangling end and a linear strand.  A direct
merging of two dangling ends would better stabilize the system but
this event  becomes more and more unlikely with time due to the
concomitant decreases of the dangling ends population.
 From this argument it follows that the time scale of network rearrangements
$\tau_s$ in Fig.~\ref{fig:avL} is determined by the scission time of
  branch points. Indeed from Fig.~\ref{fig:trimer}(b) we find that
  $t_3\simeq 10^4\simeq\tau_s$  for
 $\lambda=1.75$ and $T=1$.

\begin{figure}[!tb]
\centering
\includegraphics[width=0.99\columnwidth]{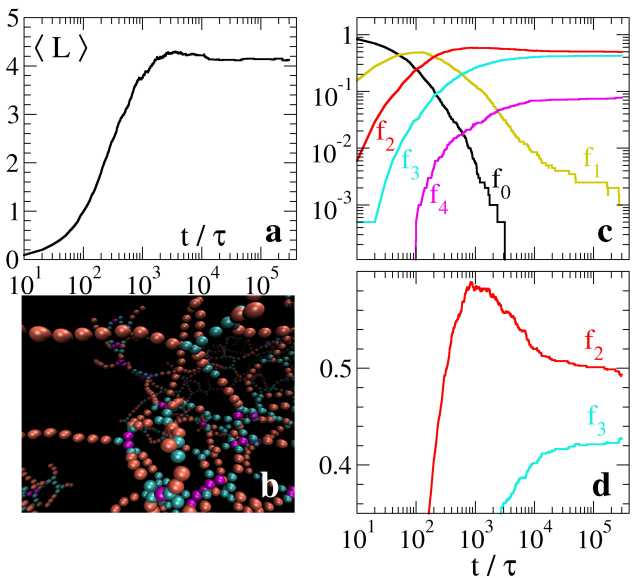}
\caption{
  Relaxation dynamics of a system with $N=2000$ particles and $\lambda= 2$ for $T=1$.
  (a)
    Average strand length $\langle L \rangle$. 
  (b)
    Final configuration of the network: gold, red, cyan, and magenta beads identify, respectively, units with $k=1,2,3,4$ contacts. 
  (c)
    Evolution of the fraction $f_k$ of units with $k$-contacts in log-log scale and (d) in linear-log scale. 
}
\label{fig:lambda2}
\end{figure}

The variety of micellar structures depends on several parameters
(solvent, salt concentration, surfactant chemistry) and includes
networks rich in branches. This situation is realized in our model by
increasing $\lambda$. For example, with $\lambda=2$ and $T=1$, the relaxation
proceeds through the formation of a metastable network with 
relatively small average strand length, as shown in
Fig.~\ref{fig:lambda2}(a) (the configuration at the end of the simulation is shown in Fig.~\ref{fig:lambda2}(b)). Again, there is a very slow convergence of the
fraction $f_1$ of units at the caps of dangling ends, see
Fig.~\ref{fig:lambda2}(c). In Fig.~\ref{fig:lambda2}(d), in linear
scale on the vertical axis, it is apparent that also the fractions $f_2$ and $f_3$
are slowly converging toward their asymptotic average values. Here, equilibrium 
thermodynamics predicts a clear dominance of branches over strands: 
in the single trimer study in fig.~\ref{fig:trimer}(a), we obtain $\Delta F_{32}\simeq -8$ 
for the nearby parameters $\lambda=1.95$ and $T=1$. Therefore, we conclude that
equilibrium is far from being reached and that the configuration reported in 
 Fig.~\ref{fig:lambda2}(b) is still a metastable one.

\begin{figure}[!tb]
\centering
\includegraphics[width=0.75\columnwidth]{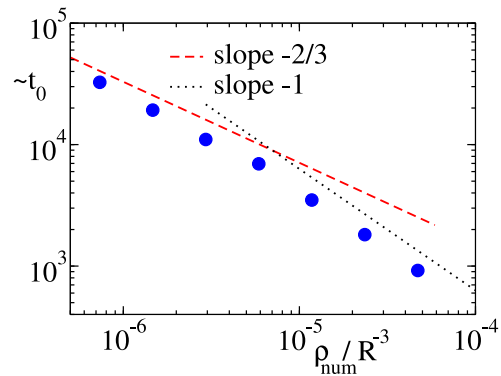}
\caption{Time scale of the network condensation as a function of the particle density.
The straight lines indicate the scalings expected at low and high densities.
 Data  refer to system sizes $N= 25, 50, 100, 200, 400, 800, 1600$, $T=1$ and $\lambda=1.8$.
 Particles are evolved in a cubic box with side equal to $324 R$ and periodic 
boundary conditions.
}
\label{fig:t0}
\end{figure}

The phenomenology  described above holds for several  particle
densities. Indeed, by changing considerably $N$ and keeping the volume
$V$ fixed, we have been able to vary  $\rho_{\rm num} = N / V$ by a
factor of $3$ (data not shown)
Yet, the scaling of the mean quantities is quite
similar to the previous cases.  In particular, the average strand
length in the initial metastable network is unaffected by the change
of density.  However, the density determines the time scale $t_0$ of
the initial network condensation. 
This is defined as the time needed for the fraction of 0-contacts
$f_0$ to go below the threshold value $f^* = 0.4$ starting from a homogeneous initial condition of isolated 
particles  $(f_0=1)$ at time $t=0$. 
By scaling arguments, at low
concentrations, diffusion leads to $t_0\sim \rho_{\rm num}^{-2/3}$. At
higher concentrations it is rather the ``ballistic'' expansion of the
network core (as in other Brownian coagulation systems~\cite{ves10})
that determines the time scale $t_0\sim \rho_{\rm num}^{-1}$.  In
Fig.~\ref{fig:t0} we report estimates of $\sim t_0$ 
(up to a constant that is irrelevant for the scaling)
which are compatible with the two regimes, see also the appendix for further details.
Note that we have verified that $f^*$ is irrelevant for the scaling behavior of $t_0$ 
with $\rho_{\rm num}$ in so far as $f^*$ is large enough to 
provide a sampling of the initial relaxation stages.

\section{Discussion}\label{sec:disc}
We have introduced a patchy particle  model to study the relaxation dynamics of living polymer networks 
driven out equilibrium by a deep thermal quench.  
A novel relevant feature of this coarse-grained  description is its ability to  bridge
the short time dynamics of its elementary motifs (dangling ends, linear strands, branches)
with the long time behavior of the mesoscopic network.

By combining large-scale simulations and analytical estimates of the
free energy barriers between  motifs,  we  rationalize the observed
slow convergence of the network to equilibrium in
terms of the  frustrated dynamics of the local motifs. In particular
we argue that the relaxation process develops through a sequence of
rearrangements that are not only energetically challenging but also
require the solution of topological constraints.
 Importantly,  such  constraints do not derive from 
static functional properties  of motifs, but emerge dynamically and 
depend  on local thermodynamic conditions.
Moreover, the  far-from-equilibrium character
of our setup  is naturally beyond the perturbative regime, where exponential
relaxation laws are expected~\cite{tur90,tur93}.

While we have used a standard temperature quench to  produce a clean
representation of the network relaxation, in practice the
considerations we have drawn should be extended to transient regimes
of mechanical or chemical origin. For instance, a passage of a
micro-bead through a portion of a viscoelastic fluid made of wormlike
micellar networks may ``scramble'' severely  the system by
generating dangling ends and by distorting the branches. This
unbalanced state then should re-equilibrate by following a relaxation
dynamics akin to the one we have  characterized here.
Such scenario is  relevant for microbeads
of diameter $D$ traveling with mean velocity $v$
at high Weissenberg number ${\rm Wi} = v / v_r$, where
$v_r = D/\tau_{\rm rel}$ is the velocity scale associated to the
relaxation over $D$ with time scale $\tau_{\rm rel}$.
Our findings suggest that a proper definition of $\tau_{\rm rel}$ should consider the 
possible emergence of power-law behaviors for strongly disturbed polymer networks.  
In fact, the relaxation times found here are of the same order of
magnitude (seconds) of those measured experimentally in viscoelastic micellar
networks~\cite{gom15,ber18}. It is thus natural to conjecture that $\tau_{\rm rel}$ 
is the time at which each microscopic motif of the network reaches, through  a
power-law decay,  its typical (equilibrium) fraction.
In this respect, it is the ending time of this scale-free aging (not to be confused with chemical aging, as a
high-temperature degradation of carbon-carbon double
bonds~\cite{chu11}) that should be considered for evaluating the
standard  Deborah and Weissenberg numbers used in
experiments~\cite{dea10,poo12}.

Our findings also raise the question on which of, e.g., self-healing rubber~\cite{cor08,gre08}, colloidal systems~\cite{del05,zac08,ang14}, or network fluids~\cite{dia17} do experience a relaxation process characterized by a slow rearrangement of local motifs. In this respect we believe that coarse-grained 
descriptions based on patchy colloids, as the one presented here, can be highly valuable.

Finally, another perspective of our work is related to the problem of the stability of ring solutions versus network structures in self-assembling systems~\cite{almarza2012,rovigatti2013}. Suitable generalizations of our model could be considered in order to bias the aggregation of linear structures towards the formation of closed rings. For instance, instead of units with two antipodal sticky spots, one could consider patchy particles that are non-axisymmetically distributed along the surface~\cite{rovigatti2013}.  
The interplay between the branching parameter $\lambda$ here introduced and the propensity to create rings is an open question that will deserve future investigations.



\appendix
\section{Numerical methods}\label{sec:app}

\paragraph*{Langevin dynamics--}
Let us consider an ensemble of $N$ patchy particles labeled by the index $i=1,\cdots,N$ and let $\alpha=1,2,3$
be an additional index that identifies the  subunits, where $\alpha=1$ is the central repulsive core   
and $\alpha=2,3$ identify the two sticky spots in Fig.~\ref{fig:sk}(a).  
In the following we denote  by $\vec{r}_{i\alpha}$  the position of the center of mass of the subunit $\alpha$ of unit $i$ and 
by $d_{i\alpha}^{j\beta} = |\vec{r}_{i\alpha} - \vec{r}_{j\beta}|$ the distance between two subunits.
The repulsive interaction between a pair of central cores of radius $R$ is modeled by a truncated and shifted Lennard-Jones potential

\begin{equation}
 U_{LJ}( D_{ij} ) = 4\epsilon \left[ \left(\frac{2R}{D_{ij}}\right)^{12}-
\left(\frac{2R}{D_{ij}}\right)^6 - \frac{1}{4}\right]  \Theta\left(x \right), 
\end{equation}

 where we have introduced the symbol $D_{ij}=d_{i1}^{j1}$ to simplify the notations. The parameter $\epsilon$ is the typical
 energy scale of the Lennard-Jones potential,  $\Theta(x)$ is the Heaviside function and $x=2^{\frac{1}{6}}(2R)  -D_{ij}$.  
 
The attractive interaction between sticky spots is described by a Gaussian potential with amplitude $A$ and range $\sigma$, namely
\begin{equation}
U_G (d_{i\alpha}^{j\beta})= - A \exp\left[-\left(\frac{d_{i\alpha}^{j\beta}}{\sigma}\right)^2\right]  \,, 
\end{equation} 
where the subindices $\alpha$ and $\beta$ are restricted to the values $(2,3)$. Moreover we set
\begin{equation}
\begin{cases}
U_{LJ}( d_{i\alpha}^{j\beta})=0  &\quad  \mbox{ if } \alpha \mbox{ or } \beta = (2,3) \\
U_G(d_{i\alpha}^{j\beta})=0      &\quad  \mbox{ if } \alpha \mbox{ or } \beta = 1
\end{cases}\quad.
\end{equation}

Subunits within each  unit evolve as a rigid body arranged as in Fig.~\ref{fig:sk}(a). 
The evolution of the system is described in terms of a Langevin equation

\begin{equation}
\label{eq:evol}
m \frac{d^2 \vec{r}_{i\alpha}}{dt^2} = -\gamma \frac{d \vec{r}_{i\alpha}}{dt} - \vec{\nabla} U (\vec{r}_{i\alpha}) + \vec{\eta}_{i\alpha}(t) \,, 
\end{equation}

where $U=U_{LJ} + U_G + ${\it constraints}  is the total potential energy function  and $m$ is the mass of each subunit.
The first term in the r.h.s. of Eq.~(\ref{eq:evol}) is a linear dissipation proportional to the friction coefficient $\gamma$, while 
$\vec{\eta}_{i\alpha}(t)$ is a Gaussian noise with zero mean and delta-correlated in time. 
 Denoting by $\eta_{i\alpha}^{(\ell)}$ with $\ell=1,2,3$ the Cartesian component of $\vec{\eta}_{i\alpha}$
in the $\ell$ direction, the noise satisfies a standard fluctuation-dissipation relation in the form
\begin{equation}
  \left\langle \eta_{i\alpha}^{(\ell)}(t)\, \eta_{j\beta}^{(\ell^\prime)}(t') \right\rangle = 2\gamma k_BT\delta_{i,j}\delta_{\alpha,\beta}\delta_{\ell,\ell^\prime}\delta(t-t') \,,
\end{equation}  

where $k_B$ is the Boltzmann constant and $T$ the temperature.
Numerical integration of Eq.~(\ref{eq:evol}) is performed with LAMMPS engine~\cite{Plimpton1995a}.
Nonequilibrium  simulations of the system relaxation process are 
realized 
according to the following protocol.
First we evolve the system described by Eq.~(\ref{eq:evol}) in the absence of the Gaussian attractive potential $(U_G=0)$
until a uniform thermalized state is reached.   Then we activate $U_G$ and we start monitoring $\langle L \rangle$ and $f_k$ as discussed
in the main text.

\paragraph*{Physical units--}
In our simulations the system temperature $T$ is measured in units of $\epsilon/k_B$, where $\epsilon$ is the amplitude
of the Lennard-Jones potential, with $k_B=\epsilon=1$, $A=40 \epsilon$ and $\sigma=0.4 R$. Distances are measured in units of monomer radii $R$.
We set $m/\gamma=\tau$, where $\tau=R\sqrt{m/\epsilon}$ is the (unitary) characteristic simulation time.
 In order to estimate the time 
$\tau$ in physical units, one can proceed as follows.
 From the Stokes formula of the friction coefficient for spherical beads with radius $R$, we have $\gamma=6\pi \eta_{sol}R$, where $\eta_{sol}$ is the viscosity of the solvent. By eliminating $m$ in the above formulas for the characteristic time, one has $\gamma=\epsilon \tau/(2 R)^2$, hence
 \[
 \tau=\frac{3 \pi\eta_{sol}(2 R)^3}{\epsilon} \simeq 2 \mu s
 \]
In the numerical evaluation we consider the nominal water viscosity at room temperature, $\eta_{sol}\simeq 1\,{\rm mPa}\,s$, and we set $T= \epsilon / k_B = 300K$ and $R = 5\,n m$.

For the large-scale simulations of this work, the system is evolved in a prism of sides $(d,d,3 d)$ with $d=108 R$, with periodic boundary conditions. Its volume corresponds to 
$3 d^3 \simeq 0.5 (\mu\,\!m)^3$. If $N=6000$ units are introduced in this volume, their volume fraction becomes
\[
\phi = \frac{N \frac 4 3 \pi R^3}{3 d^3} \simeq 0.7 \times 10^{-2}
\]

\begin{figure}[tb]
\centering
  \includegraphics[width=\columnwidth]{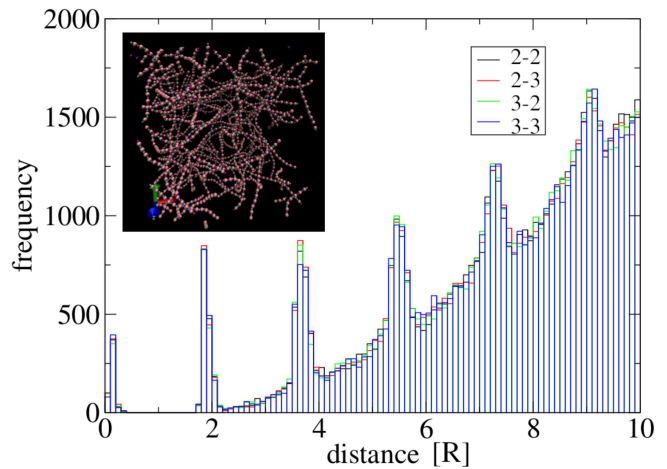}
\caption{Distribution of distances between sticky spots for a system of $N=2000$ particles with $\lambda=	1.75$, $A=40\epsilon$ and $\sigma=0.4 R$.
Particles are prepared in a homogeneous state and evolved at $T=1$ according to Eq.~(\ref{eq:evol}) in a cubic box with side equal to $54 R$ and
periodic boundary conditions.
The histogram is representative of the configuration obtained at $t=10^5 \tau$, see the inset. Black, red, green and blue bars refer respectively to 
the combination of indices $(\alpha,\beta)=\{(2,2), (2,3), (3,2), (3,3)\}$ in the set of distances $d_{i\alpha}^{j\beta}$
for $i,j=1,\cdots,N$.   Their distributions overlap very well, as expected from symmetry reasons.
Multiple peaks above the gap region derive from the regular spacing of particles in the strands.
}
\label{fig:hist_dist}
\end{figure}

\paragraph*{Contact threshold--}
Given a configuration of patchy particles identified by the set of $\vec{r}_{i\alpha}$, two distinct units $i$ and $j\neq i $  are considered in contact if 
there exists a pair of sticky spots 
$\alpha$ and $\beta$ such that $d_{i\alpha}^{j\beta}\leq  \theta$, where $\theta$ is a threshold.
 To perform an accurate sampling of merging and scission
  events in the system of particles,   $\theta$ needs to be of the order of the 
distance between bound units, in a way that typical thermal fluctuations in their positions of 
do not produce ``spurious''  events. 
 To determine a reliable value of $\theta$ for the  potential $U$ specified in the previous section, we have
computed the distributions of sticky spot distances $d_{i\alpha}^{j\beta}$ for all $i\neq j$ in generic networks, see
Fig.~\ref{fig:hist_dist} for an example. The distribution displays a clear gap in the region $[0.5 R, 1.8 R]$ separating the ensemble of bound sticky spots 
(with $d_{i\alpha}^{j\beta}\lesssim 0.5 R \simeq \sigma$) from the ensemble of unbound ones. 
Accordingly, we have chosen $\theta=R$, well inside the gap. We have verified that this threshold value is reliable 
in  the whole range of $\lambda$ and $T$ considered in this work. 
 
\paragraph*{Parallel tempering simulations--}
Free-energy differences between microstates ``2'' (strands) and ``3'' (vertices) have been computed through Langevin equilibrium simulations
with parallel tempering sampling technique~\cite{Swendsen1986,tesi1996}. This method is employed to sample accurately  the
low-temperature equilibrium distribution of the system, thereby reducing the impact of very long correlation times on
computational time.

Simulations have been performed for a system of $N=3$ interacting particles confined in 
a spherical volume with  radius $R_0=10.85 R$ and volume fraction $\phi\simeq 2\times 10^{-3}$,
which corresponds to the value of particle density of  large-scale simulations with $N=2000$.
We have verified that in this minimal setup the confining volume is sufficiently larger than the typical particle
volume, so that it does not introduce significant finite-size effects.
The free energy difference $\Delta F_{32}$ is computed via the equilibrium Boltzmann distribution 
$\Delta F_{32} = -T \log(P_3/P_2)$, where $P_3$  and $P_2$ are respectively the occupation probabilities
of states ``3" and ``2" sampled over the equilibrium Brownian dynamics.
Parallel tempering evolution is performed for a total time of $5\times 10^6\,\tau$ with a swapping period of
$50\tau$. Temperature swaps are realized between adjacent temperature states within the set of temperatures considered
in  Fig.~3(a).

\paragraph*{Scission times--}
Scission times in Figs.~3(b-c) have been computed for a system of   $N=3$ interacting particles confined in 
a spherical volume with  radius $R_0=10.85 R$ and evolving according to Eq.~(\ref{eq:evol}).
For each value of temperature $T$, particles were initially prepared in a ``3'' configuration (Fig.~3(b))
and in a ``2'' configuration (Fig.~3(c)) and evolved until a scission event to a dimer-plus-monomer state (configuration ``1'')
 was observed. For each $T$, the average scission time has been computed by averaging over a sample of 10
independent scission events.

\paragraph*{Condensation  time scales--}
 

\begin{figure}[tb]
\centering
  \includegraphics[width=\columnwidth]{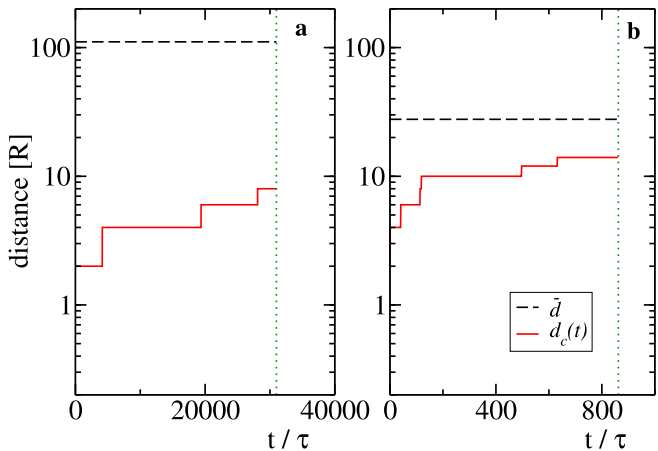}
\caption{Evolution of the maximal cluster size $d_c(t)$ (red solid lines) for $N=25$ (panel (a))  and $N=1600$ (panel (b)). 
Black dashed horizontal lines indicate the values of the typical particle distance $\bar{d}$ while green vertical lines
identify the time at  which $f_0$ reaches the threshold value $f^*$. 
}
\label{fig:max_clust}
\end{figure}

Diffusive and ballistic  regimes are further analyzed in Fig.~\ref{fig:max_clust}, where we compare the average particle distance $\bar{d}=\rho_{num}^{-1/3}$ with
the typical size $d_c(t)$ of the biggest cluster of particles found in the system at time $t$. 
We define $d_c(t)=2RN_c(t)$, where $N_c(t)$ is the total number of particles in the cluster. 
 We focus on two sizes, namely 
$N=25$ (Fig.~\ref{fig:max_clust} (a)) and $N=1600$ (Fig.~\ref{fig:max_clust} (b)), which belong respectively to the diffusive and ballistic
 condensation regime.
For $N=25$ we find $\bar{d}\gg d_c(t)$ during the whole period in which $f_0>f^*$  while the case $N=1600$
 displays a quick growth of $d_c(t)$ to values which are of the same order of $\bar{d}$ thus suppressing the diffusive 
 scaling of $t_0$ ~\cite{ves10}.


 \section*{Conflicts of interest}
 There are no conflicts to declare.

\section*{Acknowledgments}
We thank Alberto Rosso and Riccardo Sanson for useful discussions.
We acknowledge support from Progetto di Ricerca Dipartimentale BIRD173122/17 of the University of Padova. Part of our simulations were performed in the CloudVeneto platform.





%

\end{document}